# *SaVi*: satellite constellation visualization

## Lloyd Wood

*Research Fellow, Centre for Communication Systems Research at the University of Surrey, e-mail: L.Wood@surrey.ac.uk*

**Abstract**

*SaVi*, a program for visualizing satellite orbits, movement, and coverage, is maintained at the University of Surrey. This tool has been used for research in academic papers, and by industry companies designing and intending to deploy satellite constellations. It has also proven useful for demonstrating aspects of satellite constellations and their geometry, coverage and movement for educational and teaching purposes. *SaVi* is introduced and described briefly here.

## 1. Introduction

*SaVi*, the Satellite Visualization tool [1], is a computer program for visualizing and animating the movement of satellites and their coverage. *SaVi* was originally developed by Worfolk *et al.* at the Geometry Center at the University of Minnesota, but became homeless when that was closed due to lack of ongoing funding. Maintenance of the software was taken over by Lloyd Wood, who had found *SaVi* useful during his doctoral work on satellite constellations. *SaVi* has been maintained at the University of Surrey since then.

## 2. Technical Approach

*SaVi* exists as a standalone program that can also be run as a 'module' that interfaces with and controls the *Geomview* program [2]. *Geomview* is a general-purpose rendering program useful to mathematicians; *SaVi* leverages *Geomview* for simple three-dimensional (3D) rendering and OpenGL texturemapping, while ignoring *Geomview*'s ability to render higher dimensions of interest to mathematicians.

*SaVi* is implemented as a satellite orbit simulator, written in ANSI C, which is driven by commands added to the higher-level Tool Command Language (Tcl). This two-pronged approach allows *SaVi* to be scriptable. Simple, short, Tcl scripts generating satellite constellations and driving the underlying simulator are written in a similar manner to the scripts of the network simulator *ns-2*, which also relies on Tcl. Many scripts simulating, illustrating and animating proposed and existing satellite constellations are included with *SaVi*.

*SaVi*'s user interface is presented in Tcl's Toolkit, Tk, which complements Tcl and allows for relatively straightforward creation of a graphical dialog- and window-driven system [fig. 1]. Seeing and animating a complex satellite constellation is as simple as clicking the Constellations menu and selecting, say, the *Iridium* system to run the associated script [fig. 2].

As *SaVi* relies only on Tcl/Tk and standard Unix POSIX libraries, with continued maintenance it remains portable across a wide range of Unix-compatible systems, including Linux, FreeBSD, and Mac OS X. It comes as an easily-installable Debian package for Ubuntu users.

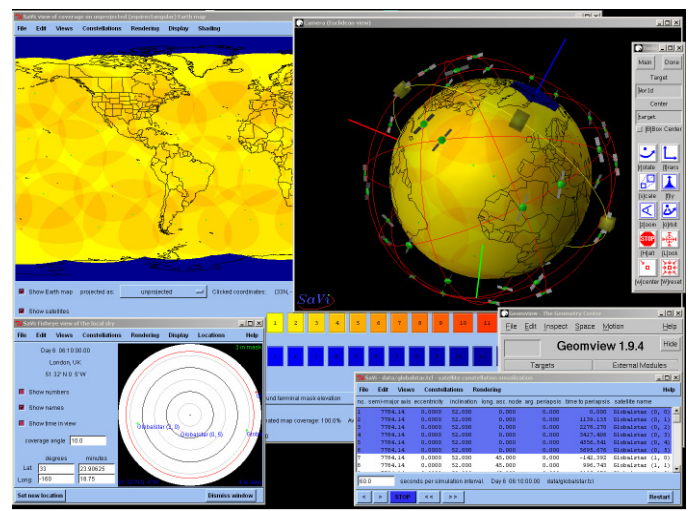

**Fig. 1:** *SaVi* user interface showing *Globalstar* simulation, with coverage, fisheye and 3D view from *Geomview*

*SaVi* can also be run under Microsoft Windows, using Cygwin or a virtualisation environment such as VirtualBox.

As a popular community-driven effort. *SaVi* is in the top 1% of projects on the SourceForge site for open and free software. *SaVi*'s portability and popularity is maintained by users reporting bugs and requesting features, or providing fixes for problems encountered with new compilers or with specific platforms. As a result, after over fifteen years of life, *SaVi* remains compatible with modern systems.

*SaVi* shows satellite coverage areas on a number of different map projections. A fisheye view of the sky is also available to examine how satellites pass over different points on the Earth. *SaVi* shows satellite coverage as either minimum elevation angle or as half-angle beamwidth, and indicates how that coverage moves over time. Graphical output can be recorded and saved. Satellite and constellation properties can be edited.

Multiple spotbeams on a satellite, communication channel properties, and precise orbital motion with complex precession are not yet simulated; the University has other, custom, tools for simulating these in far more detail.

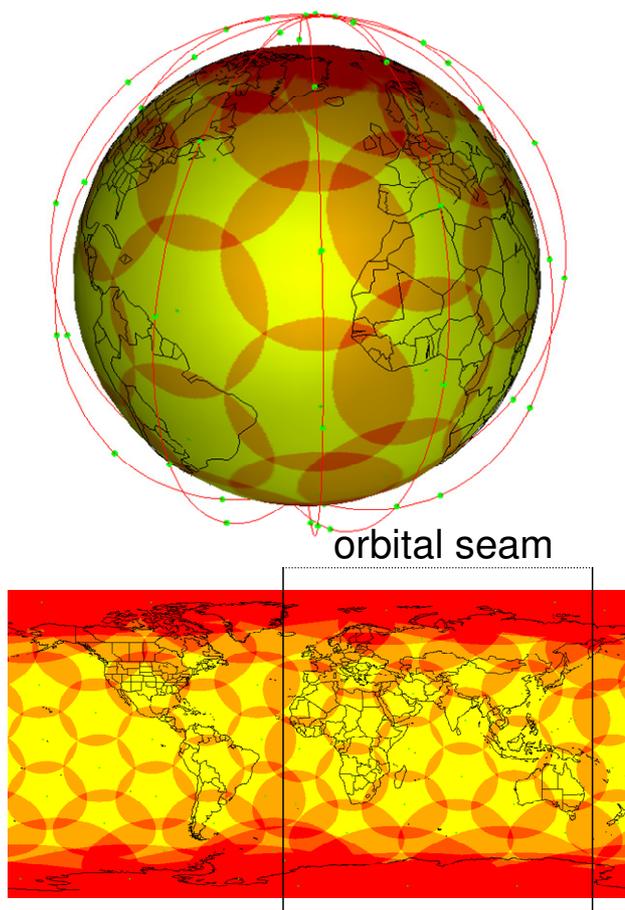
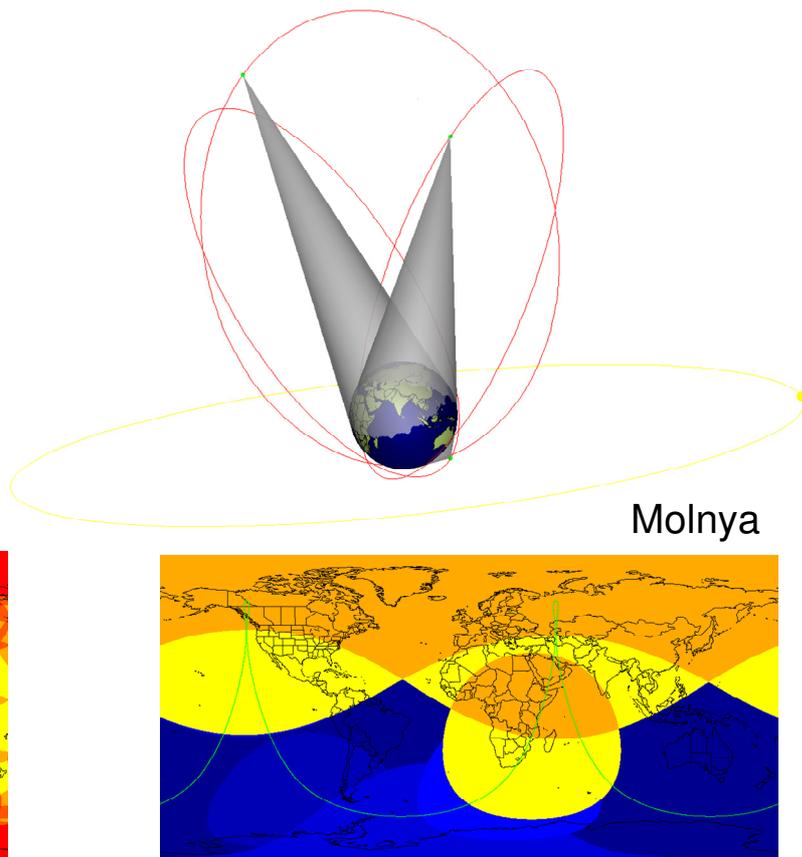

**Fig 2 – The *Iridium* constellation simulated in *SaVi***

**Fig 3 – High-latitude *Molnya* constellation compared to geostationary ring in *SaVi*, showing repeating groundtrack and movement over time**

## 3. Results

Since writing constellation scripts that shipped with *SaVi* 1.0 and adopting *SaVi*, Lloyd has added many new features, including many more scripts, new map projections, resizable coverage and fisheye displays, viewing of high-diversity constellations and coverage movement over time, moving coverage texturemaps, constellation generation tools, and an educational help system that explains each constellation.

While *SaVi* lacks the large number of features present in far richer commercial offerings such as the Satellite Toolkit (STK), it is entirely free to use, which makes *SaVi* immediately attractive to the academic community. Its output has appeared in over twenty research papers and articles [3].

*SaVi* enables quick and easy explanation of the basic features of orbital motion and satellite geometry. This can include showing the differences between rosette and star constellations, by contrasting the seamed *Iridium* and seamless *Globalst*ar systems, comparing diversity, overlapping coverage, and the large number of satellites seen in the sky for navigation constellations, or demonstrating repeating-groundtrack designs such as *Molnya* [fig. 3]. *SaVi* has been found useful for teaching purposes at the MSc level and on short courses at a number of institutions, such as the International Space University and SUPAERO, when constellations and orbital movement of non-geostationary systems must be explained.

*SaVi* has also been picked up by industry companies designing satellite constellations for communication, and has been used publically to illustrate the designs of their systems [4, 5].

## 4. Summary of the work, potential impact and conclusion

Although *SaVi* only provides a relatively simple degree of satellite simulation functionality when compared to more full-featured commercial packages, its open codebase and contributions from around the world have led to a long-lived, robust, portable, cross-platform tool that has attracted a wide degree of interest. *SaVi* appears to have gained a useful educational role in introducing and explaining the properties of satellite constellations.

### References

[1] L. Wood, P. Worfolk, *et al.*, *SaVi* 1.4.5, Sourceforge project site, April 2011. http://savi.sf.net/

[2] *Geomview* 1.9.4, Sourceforge project site, August 2007. http://www.geomview.org/

[3] L. Wood, Examples of use of *SaVi*, list of papers and articles, page last retrieved May 2011. http://savi.sf.net/lw/examples.html

[4] D. Kohn, multimedia gallery, Teledesic LLC, 1997-2002.

[5] G. Wyler, presentation to the Pacific Telecommunications Council Conference (PTC '09) on O3b Networks, January 2009.